\documentclass[10pt,twocolumn]{article}

\usepackage{amsmath}
\usepackage{amssymb}
\usepackage{booktabs}
\usepackage{multirow}
\usepackage{array}
\usepackage{xspace}
\usepackage[margin=0.72in]{geometry}
\usepackage{microtype}
\usepackage{url}
\usepackage[hidelinks]{hyperref}
\usepackage{tikz}
\usetikzlibrary{arrows.meta,positioning,fit,shapes.geometric}
\hypersetup{
  pdftitle={Chiral Analysis of Smart Contracts: Detecting Vulnerabilities from Relational Inconsistencies Across Business Paths},
  pdfauthor={Yue Xue},
  pdfkeywords={smart contracts, vulnerability detection, program analysis, inconsistency detection, large language models}
}

\microtypesetup{expansion=false}
\newcommand{\Description}[1]{}

\newcommand{\tool}{ChiralDetector\xspace}

\newcommand{\sem}[1]{\lbrack\!\lbrack #1 \rbrack\!\rbrack}

\begin{document}

\title{Chiral Analysis of Smart Contracts: Detecting Vulnerabilities from Relational Inconsistencies Across Business Paths}

\author{Yue Xue\\Independent Researcher}
\date{}

\maketitle

\begin{abstract}
Smart-contract vulnerabilities often arise not from a single obviously unsafe statement, but from inconsistencies between business operations that should correspond to one another. Examples include single and batch entry points, direct and adapter-based flows, quote and execution paths, and inverse operations such as buy and sell. Existing analyzers are effective for many local syntactic and data-flow patterns, but they have limited support for vulnerabilities whose oracle is relational: whether two semantically paired paths preserve the same guards, state transitions, value flows, and failure behavior.

This paper introduces \emph{chiral analysis}, a relational model for detecting smart-contract vulnerabilities across corresponding business paths. The key insight is to treat a pair of paths as an implicit specification for each other. We formalize a chiral relation as a static analogue of a metamorphic relation, derive relation-induced obligations over guards, actors, state, value, ordering, failure behavior, and external interactions, and report a vulnerability when a violated obligation has security impact. We implement this idea in \tool, a prototype pipeline that extracts Solidity business paths, ranks candidate path pairs using static facts, applies LLM-based semantic filtering and detection, and validates/deduplicates resulting findings.

In a preliminary evaluation on the Phi protocol, \tool reduced 3,217 statically ranked path pairs to 1,643 semantic candidates, produced 101 deduplicated finding groups, and retained 44 strict-validator positives that manually collapsed to 13 effective unique issues. These findings include high-impact relation violations such as cross-art Merkle proof reuse, fee unit mismatches, public state-tracking helpers, and refund propagation gaps. The results suggest that chiral analysis can expose bug classes that are difficult to express as single-function rules, while providing a structured way to control LLM cost and validator precision.
\end{abstract}


\section{Introduction}

Smart-contract security analysis increasingly needs to reason about business logic rather than isolated unsafe statements. A decentralized application often exposes several paths for the same underlying operation: a direct claim and an adapter claim, a single trade and a batch trade, a view quote and an execution path, or a buy path and a sell path. These paths are not necessarily syntactically similar, and they may cross different contracts or helper functions. However, they often share a business invariant: the same actor should be authorized, the same fee should be charged, the same state should be consumed, or the reverse path should undo a compatible state transition.

Existing smart-contract analyzers are strong at local vulnerability templates, taint-style dependencies, and syntactic or intermediate-representation facts~\cite{tsankov2018securify,tikhomirov2018smartcheck,feist2019slither}. These techniques are valuable, but many business-logic bugs do not have a single local syntactic signature. For example, whether an adapter traps a refund depends on comparing the adapter path with the core path. Whether a batch operation preserves single-operation semantics depends on comparing per-item checks, value propagation, and failure behavior. Whether a quote is safe depends on comparing it against the execution path it predicts. In these cases, the missing oracle is relational: the bug exists because two corresponding paths disagree.

Software-engineering research has long used inconsistency as a source of bug oracles. Rule-mining systems infer likely programming rules from repeated code facts and flag violations~\cite{engler2001bugs,li2005prminer}. Clone and porting-bug detectors compare corresponding code regions to detect suspicious differences~\cite{li2006cpminer,ahmadi2021fics}. Metamorphic testing defines relations between multiple executions when a single exact oracle is unavailable~\cite{segura2016metamorphic}. These ideas suggest a natural framing for smart contracts: instead of asking whether one function is locally unsafe, we can ask whether two business paths that should correspond violate their induced relational obligations.

This paper proposes \emph{chiral analysis}. We call two paths \emph{chiral} when they are semantically paired but not required to be syntactically identical. The term is intended to capture business operations that are mirror-like, alternative, wrapper/core, single/batch, lifecycle-related, or preview/execution-related. A chiral pair induces obligations over semantic dimensions such as guards, actors, state updates, asset flows, ordering, failure behavior, and external dependencies. A chiral vulnerability is a violation of such an obligation with a feasible security or protocol impact.

\tool operationalizes this idea for Solidity projects. It first extracts externally reachable business paths from a call graph and summarizes each path using static facts over storage reads/writes, value transfers, modifiers, visibility, external calls, and entry types. It then ranks path pairs by shared state and call-footprint overlap to avoid exhaustive semantic comparison. The remaining pairs are passed through a low-cost semantic filter and a relation-aware detector. Finally, a strict validator attempts to falsify each finding against the source code, and a deduplication step merges reports that share the same root cause and likely fix.

Our preliminary results on the Phi protocol show that chiral analysis can find concrete inconsistencies that are not naturally expressed as single-function detectors. From 3,217 static candidate pairs, \tool kept or marked 1,643 pairs for semantic detection. The detector produced 201 cleaned finding/suspicious candidates, which were locally deduplicated to 101 groups. A strict validator retained 44 positives, and manual root-cause deduplication reduced these to 13 effective unique issues. These include cross-art Merkle proof reuse, fee-unit confusion, unauthorized public tracking helpers, refund propagation errors, and single/batch guard asymmetries.

This paper makes the following contributions:
\begin{itemize}
  \item We introduce chiral analysis, a relational vulnerability model for smart contracts that treats corresponding business paths as implicit specifications for each other.
  \item We formalize chiral relations as static analogues of metamorphic relations and define relation-induced obligations over seven semantic dimensions.
  \item We design \tool, a path-centric detection pipeline that combines static pair reduction, LLM semantic judgment, strict validation, and root-cause deduplication.
  \item We provide a preliminary evaluation on a real Solidity protocol, showing that the model identifies 13 manually deduplicated issues from relation-level inconsistencies.
\end{itemize}

\subsection{Scope and Non-Goals}

This draft is written as a technical report rather than a compact conference submission. Our immediate goal is to make the problem formulation, detection logic, and early empirical evidence explicit enough for public discussion and replication. We therefore include design choices, negative cases, validation logic, and future benchmark plans that would normally be compressed or moved to an appendix in an eight-to-ten-page venue format.

The scope of the current prototype is Solidity source analysis. We focus on vulnerabilities whose evidence depends on comparing two or more business paths. We do not aim to replace local detectors for reentrancy, unchecked calls, arithmetic mistakes, or access-control bugs when those bugs have clear single-function signatures. Instead, chiral analysis targets the cases where a local detector lacks an oracle: the suspicious behavior is only suspicious because a related path handles the same business object differently.

We also do not claim that every chiral mismatch is exploitable. Some mismatches are documentation bugs, low-severity integration hazards, or protocol-design inconsistencies. The model therefore separates four levels: candidate relation, suspicious mismatch, validated vulnerability, and root-level finding. This separation is central to the method because relation mining is intentionally recall-oriented while final reporting must be precision-oriented.

\section{Motivating Example}

This section motivates chiral analysis using representative examples from our preliminary Phi study. The purpose is not to claim that every inconsistency is exploitable, but to show why the relevant oracle lives across paths.

\subsection{Direct and Batch Claim Paths}

Consider a protocol that exposes both a direct claim and a batch claim. A local analyzer may inspect each function independently and find no obvious unsafe primitive. However, the two paths induce a single/batch relation: a batch claim should behave like repeated direct claims under the protocol's declared atomicity and refund model. If the direct path recomputes a mint fee and refunds excess payment, while the batch dispatcher forwards per-item values in a way that leaves excess ETH in the factory, the bug is visible only by comparing the two business paths. The violated obligation is a value-flow obligation: equivalent claim operations should preserve the same payer, charged amount, and refund recipient.

\subsection{Quote and Execution Paths}

Quote functions are another common source of relational oracles. A view path often predicts the result of an execution path, and users or integrators rely on that prediction to make economic decisions. If a boundary case is handled differently in the quote path and the execution path, the local code may still look intentional. The problem emerges when the preview/execution relation is applied: both paths should use compatible price sources, fees, rounding directions, and boundary conditions. A mismatch is not always a high-severity vulnerability, but it is a protocol-relevant inconsistency that should be reported and triaged.

\subsection{Why This Is Not Clone Detection}

The paired paths in these examples are not necessarily clones. They may have different entry points, wrappers, helper calls, or path lengths. They may also be semantically opposite rather than similar, as in buy/sell or mint/burn. Chiral analysis therefore cannot rely only on syntactic similarity. It needs a program representation that preserves call-graph structure and semantic facts, and a relation language that can express why two different-looking paths should nevertheless constrain each other.

\subsection{A Concrete Relation Failure}

The clearest way to understand the model is to view a pair as a small relational specification. Suppose a factory exposes a direct claim path and an NFT adapter exposes a wrapper claim path. The direct path receives ETH from the user, computes the required minting fee, mints the asset, and refunds excess value to the caller. The wrapper path receives ETH from the same user-facing call, forwards value to the factory, and then relies on the factory's refund behavior. A wrapper/core relation says that the adapter should preserve the identities of payer, receiver, and refund recipient unless the protocol intentionally documents a different custody model.

The violation is not ``the adapter calls the factory'' or ``the factory refunds ETH''. Each local behavior can be reasonable in isolation. The violation is that the composition changes the sink of excess value: the refund may return to an intermediate contract rather than to the original user-facing payer. In the notation of this paper, the actor obligation fails because the refund actor in $\sem{p}$ is not preserved by the wrapper transformation $T_I,T_O$ in $\sem{q}$. This kind of finding is hard to express as a single syntactic rule, but natural as a failed chiral obligation.

\subsection{A Non-Bug Chiral Mismatch}

Chiral analysis must also tolerate benign asymmetry. A batch path may intentionally be all-or-nothing while the corresponding single path succeeds or fails one item at a time. A preview function may intentionally be conservative rather than exactly equal to execution. A privileged rescue path may bypass ordinary lifecycle checks because its purpose is emergency recovery. These cases should not be reported as vulnerabilities merely because two paths differ.

For this reason, every relation type carries its own expected transformation rather than a universal equality check. A single/batch relation requires an explicit atomicity model. A preview/execute relation allows conservative rounding. A lifecycle relation allows phase-specific guards, but requires that later phases not invalidate the state assumptions created by earlier phases. This typed interpretation is the main difference between chiral analysis and a naive inconsistency detector.

\section{Chiral Vulnerability Model}

\subsection{Program and Path Semantics}

We model a Solidity project as a labeled transition system:
\begin{equation}
M = (\Sigma, I, O, \rightarrow),
\end{equation}
where $\Sigma$ is the set of blockchain states, $I$ is the set of transaction inputs, $O$ is the set of observable outputs, and $\rightarrow$ is the execution relation. The state includes contract storage, balances, block context, and in-scope external contract state. Inputs include caller, value, calldata, signatures, and environmental assumptions. Outputs include return values, reverts, logs, ETH/token transfers, and external calls.

A business path $p$ is a semantically meaningful call-graph slice:
\begin{equation}
p = (entry, nodes, edges, cond, span).
\end{equation}
The path may be a single function, an entry-to-leaf chain, a branch segment, a wrapper-to-core path, or a small path bundle. Each path denotes a partial transition:
\begin{equation}
\sem{p}: \Sigma \times I \rightharpoonup \Sigma \times O.
\end{equation}
The transition is partial because the path may revert or be infeasible under a given input.

For static detection, \tool approximates a path by a semantic fact summary:
\begin{equation}
\Phi(p) = (A, G, R, W, \Delta S, \Delta B, X, L, Ord),
\end{equation}
where $A$ captures actors and roles; $G$ captures guards and preconditions; $R$ and $W$ are storage read/write sets; $\Delta S$ and $\Delta B$ approximate state and asset transitions; $X$ captures external calls; $L$ captures logs, errors, and reverts; and $Ord$ captures ordering constraints such as check-effect-interaction structure.

The summary is deliberately not a full formal semantics of the EVM. Full equivalence checking over smart-contract paths is too expensive and often unnecessary for audit triage. Instead, $\Phi(p)$ is a relation-oriented abstraction: it preserves the facts that commonly define business invariants. For example, a path summary keeps whether a state flag is consumed, whether an array index is aligned with another array, whether ETH is refunded to a caller or to an intermediate contract, and whether a non-reentrant guard is present before an external interaction. These facts are sufficient to ask whether two paths obey an expected relation even when the exact arithmetic or all path conditions are not fully solved.

\subsection{Relation Granularity}

The unit of a chiral relation is a business path, not necessarily a function. This choice matters because smart contracts frequently distribute one operation across wrappers, internal helpers, library calls, and cross-contract dispatchers. A function-level detector can miss a wrapper/core relation because the wrapper has little logic and the core function looks locally correct. Conversely, enumerating every subpath is infeasible because the number of possible segments grows rapidly with call-graph branching.

We use three practical granularities. First, a single-function path is retained when the function modifies state, transfers value, affects authorization, or serves as a retained view/quote for a state-changing path. Second, a straight-line entry-to-leaf path is retained as a complete business flow when intermediate calls do not branch into different function calls. Third, a fork segment is retained when a function-call branch introduces multiple downstream paths. The third case is important because many relation bugs arise at the boundary of a branch: one branch applies a check or settlement step and another branch omits it.

\subsection{Chiral Relations}

A chiral relation is a static relational oracle over two business paths:
\begin{equation}
R_\chi = (p, q, \tau, Pre, T_I, T_\Sigma, T_O, Obl).
\end{equation}
Here $p$ and $q$ are corresponding business paths, $\tau$ is the relation type, $Pre$ is the precondition under which the relation is expected to hold, $T_I$ maps inputs for one side to corresponding inputs for the other side, $T_\Sigma$ relates initial states, $T_O$ defines the expected relation over final states and outputs, and $Obl$ is a set of static obligations that approximate $T_O$.

This definition is intentionally close to metamorphic testing, but its usage is static and security-oriented. In metamorphic testing, a source input is transformed into a follow-up input and the two executions should satisfy an output relation. In chiral analysis, a source path is related to a follow-up path, and the two business operations should satisfy a relation over guards, state, value, and observable behavior. The relation can be equality-like, inverse-like, conservative, or wrapper-preserving depending on $\tau$.

The relation also has a confidence level. Some pairs are explicit: names, documentation, or direct dispatch show that they implement the same operation. Other pairs are implicit: they touch the same storage variables and participate in the same lifecycle but have no naming cue. The current prototype treats relation discovery as a candidate generation problem, not a proof obligation. A candidate pair may be rejected later if the validator cannot justify why the two paths should constrain each other.

Table~\ref{tab:relation-types} summarizes the relation types used by \tool. The table is intentionally typed: a single/batch pair should not be checked with the same equality rule as an inverse buy/sell pair or a wrapper/core pair.

\begin{table*}[t]
  \caption{Chiral relation types and their induced obligations.}
  \label{tab:relation-types}
  \centering
  \small
  \begin{tabular}{p{0.17\linewidth}p{0.27\linewidth}p{0.46\linewidth}}
    \toprule
    Relation type & Typical shape & Example obligations \\
    \midrule
    Similar flow & Alternative ways to perform the same action & Equivalent authorization, eligibility, nonce/deadline checks, consumed flags, fees, rewards, events \\
    Inverse flow & Opposite operations over the same state space & Inverse or conservation-preserving state/value deltas, compatible fees, checkpoints, cooldowns, cleanup \\
    Single/batch & Batch path corresponds to repeated single path & Per-item guard parity, array alignment, per-item state writes, explicit failure atomicity, equivalent settlement \\
    Wrapper/core & Adapter, router, or proxy exposes a core path & Caller/payer/receiver/refund preservation, unit preservation, no weakened checks, no trapped dust/rewards \\
    Branch symmetric & Symmetric branches inside a function family & Correct paired variables, caps, prices, decimals, storage keys, events, external calls \\
    Lifecycle & Create/update/delete or request/execute/claim phases & Compatible creation and update constraints, cleanup of introduced state, valid state-machine transitions \\
    Shared settlement & Multiple entries converge on a helper or accumulator & Consistent settlement timing, indexes, reserves, reward debt, claim flags, and canonical helper usage \\
    Preview/execute & View path predicts an execution path & Compatible price source, fee model, rounding, boundary handling, conservative preview semantics \\
    \bottomrule
  \end{tabular}
\end{table*}

\subsection{Relation-Induced Obligations}

An obligation is a checkable approximation of a chiral relation:
\begin{equation}
o = (d, f_p, rel, f_q, pol, sev),
\end{equation}
where $d$ is a semantic dimension, $f_p$ and $f_q$ are facts extracted from the two paths, $rel$ is the expected relation between the facts, $pol$ indicates whether the observed relation satisfies or violates the obligation, and $sev$ is a severity hint.

We use seven obligation dimensions. \emph{Guard} obligations compare modifiers, require checks, pause checks, deadline checks, nonce checks, eligibility checks, and signature requirements. \emph{Actor} obligations compare caller, signer, payer, receiver, beneficiary, and refund identities. \emph{State} obligations compare read/write sets, mapping keys, flags, counters, checkpoints, and state deltas. \emph{Value} obligations compare units, fees, rewards, transfers, mint/burn behavior, refunds, and dust handling. \emph{Order} obligations compare settlement timing and check-effect-interaction order. \emph{Failure} obligations compare revert behavior, partial success semantics, errors, and events. \emph{External} obligations compare oracle usage, callbacks, cross-contract calls, upgrade/config dependencies, and trust boundaries.

\begin{table*}[t]
  \caption{Obligation dimensions used to turn a chiral relation into checkable evidence.}
  \label{tab:obligations}
  \centering
  \small
  \begin{tabular}{p{0.12\linewidth}p{0.29\linewidth}p{0.48\linewidth}}
    \toprule
    Dimension & Compared facts & Typical violation question \\
    \midrule
    Guard & Modifiers, require checks, pause/deadline/nonce/signature checks & Does one side admit an input that the paired side rejects for the same business object? \\
    Actor & Caller, signer, payer, receiver, beneficiary, refund sink & Does a wrapper, batch dispatcher, or inverse path silently change who pays, receives, or is authorized? \\
    State & Read/write sets, mapping keys, flags, counters, checkpoints & Does one path skip, delay, or corrupt a state transition required by the paired path? \\
    Value & Units, fees, rewards, refunds, mint/burn, transfers & Are economically equivalent paths using incompatible units, fee bases, rounding, or settlement recipients? \\
    Order & Check/effect/interaction order, settlement timing & Does one side expose a race, callback, or stale-state window absent from the paired side? \\
    Failure & Revert behavior, partial success, events, custom errors & Does a batch or wrapper path hide failures, emit incompatible events, or violate declared atomicity? \\
    External & Oracles, callbacks, cross-contract calls, upgrade/config dependencies & Does one side rely on a weaker trust boundary or stale dependency than its pair? \\
    \bottomrule
  \end{tabular}
\end{table*}

\subsection{From Mismatch to Vulnerability}

A mismatch alone is not a vulnerability. A reportable chiral vulnerability requires a violated obligation and a witness with protocol-relevant impact:
\begin{equation}
v = (\chi, o, w, Impact).
\end{equation}
The witness has the form:
\begin{equation}
w = (\sigma, i_p, i_q, pc, evidence),
\end{equation}
where $pc$ is the path condition and $evidence$ links the violation to source locations. Informally, a finding is valid when $Pre(\sigma,i_p,i_q)$ holds, both path executions are feasible or their failure behavior is comparable, the expected output/state relation $T_O$ is violated, and the violation can cause loss, unauthorized state change, denial of service, griefing, accounting error, or another protocol-relevant impact.

This distinction is important for precision. Chiral analysis produces candidates, suspicious mismatches, validated vulnerabilities, and root-level findings as separate artifacts. This separation follows prior SE practice: inconsistency signals are useful bug oracles, but not every inconsistency is a confirmed bug.

\subsection{Taxonomy of Chiral Vulnerabilities}

The vulnerability taxonomy is two-dimensional. The first axis is the relation type $\tau$, which describes why two paths should correspond. The second axis is the violated obligation dimension $d$, which describes what kind of security-relevant mismatch occurred. A reportable class can therefore be named as a pair such as single/batch value-flow violation, wrapper/core actor-binding violation, inverse-flow state-delta violation, or preview/execute fee-model violation.

This two-axis structure avoids an overly long flat list of bug classes. It also helps reviewers and auditors understand generality. For example, ``refund trapped in adapter'' and ``reward trapped in router'' differ in protocol details but share a wrapper/core actor/value violation. ``Batch claim does not refund excess value'' and ``batch trade delays lock validation'' differ in impact but share a single/batch relation. This grouping supports both detector design and root-cause deduplication.

We use three severity-facing labels in the current prototype. A \emph{finding} has a concrete violated obligation, source evidence, and plausible impact. A \emph{suspicious} candidate has a real relation and mismatch but incomplete impact evidence. A \emph{discussion} item identifies a design inconsistency that may be intentional, low severity, or dependent on protocol policy. The validator can reject any of these labels if the relation itself is unjustified or if the claimed impact is contradicted by the source code.

\section{Method}

\subsection{Overview}

\begin{figure*}[t]
  \centering
  \begin{tikzpicture}[
    node distance=1.0cm and 1.2cm,
    box/.style={draw, rounded corners=2pt, align=center, minimum width=2.7cm, minimum height=0.9cm, font=\small},
    wide/.style={draw, rounded corners=2pt, align=center, minimum width=3.4cm, minimum height=0.9cm, font=\small},
    arr/.style={-Latex, thick}
  ]
    \node[box] (src) {Solidity\\project};
    \node[wide, right=of src] (paths) {Call graph and\\business paths};
    \node[wide, right=of paths] (facts) {Path fact\\summaries};
    \node[wide, right=of facts] (pairs) {Static pair\\ranking};
    \node[wide, below=of pairs] (sem) {LLM semantic\\filter};
    \node[wide, left=of sem] (detect) {Relation-aware\\detection};
    \node[wide, left=of detect] (valid) {Strict validation\\and dedup};
    \node[box, left=of valid] (out) {Root-level\\findings};
    \draw[arr] (src) -- (paths);
    \draw[arr] (paths) -- (facts);
    \draw[arr] (facts) -- (pairs);
    \draw[arr] (pairs) -- (sem);
    \draw[arr] (sem) -- (detect);
    \draw[arr] (detect) -- (valid);
    \draw[arr] (valid) -- (out);
  \end{tikzpicture}
  \caption{\tool pipeline. Static analysis reduces the pair space before LLM-based semantic judgment; validation and deduplication separate candidate mismatches from root-level findings.}
  \Description{A left-to-right and top-to-bottom pipeline from Solidity project, call graph and business paths, path fact summaries, static pair ranking, LLM semantic filter, relation-aware detection, strict validation and deduplication, to root-level findings.}
  \label{fig:pipeline}
\end{figure*}
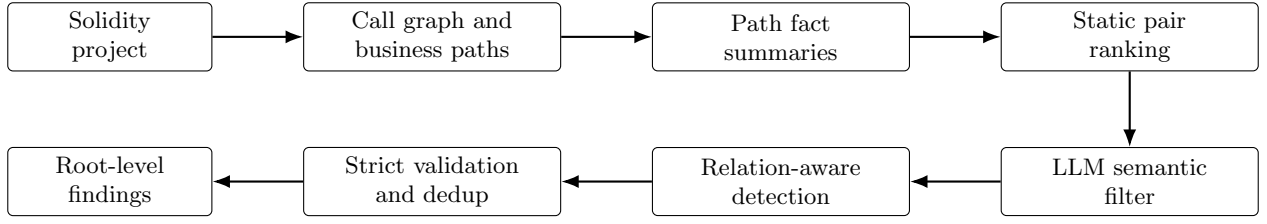

\tool detects chiral vulnerabilities in four stages: path extraction, candidate reduction, semantic detection, and validation/deduplication. Figure~\ref{fig:pipeline} shows the pipeline. The design goal is recall-first candidate discovery followed by precision-oriented validation. This division is necessary because exhaustive pairwise LLM judgment over all paths is costly, while purely static rules are too brittle for semantically opposite or wrapper-shaped relations.

The pipeline is designed around a simple observation: static analysis is cheap enough to run broadly, while LLM semantic judgment is expressive enough to recognize business relations but too costly to apply without structure. \tool therefore uses static analysis to preserve candidate recall and reduce the pair space, then uses the LLM only on concrete path pairs with explicit source evidence. The final validator is deliberately stricter than the detector, because the detector's job is to surface possible relational bugs while the validator's job is to protect the final report from overclaiming.

\subsection{End-to-End Algorithm}

The end-to-end algorithm is shown in Algorithm~\ref{alg:pipeline}. The input is a Solidity project and a budget configuration. The output is a set of root-level findings with duplicate candidate identifiers retained for auditability.

\begin{table*}[t]
  \caption{High-level \tool algorithm.}
  \label{alg:pipeline}
  \centering
  \small
  \begin{tabular}{p{0.96\linewidth}}
    \toprule
    \textbf{Input:} Solidity project $P$, relation budget $B$, validator budget $V$ \\
    \textbf{Output:} root-level findings $F$ \\
    \midrule
    1. Parse contracts and build a project-level call graph $G$. \\
    2. Extract retained business paths $\mathcal{P}$ from external entries, single functions, straight-line chains, and fork segments. \\
    3. Compute path summaries $\Phi(p)$ for each $p \in \mathcal{P}$. \\
    4. Generate candidate pairs $\mathcal{C}=\{\langle p,q\rangle \mid p,q\in\mathcal{P}, p\neq q\}$ and score them using static relation features. \\
    5. Select the top-ranked or budget-feasible pairs $\mathcal{C}_B$ without deleting uncertain high-recall signals too early. \\
    6. Ask the semantic filter to classify whether each pair in $\mathcal{C}_B$ forms a plausible chiral relation and to assign relation type $\tau$. \\
    7. For kept pairs, ask the detector to identify violated obligations, evidence, and impact. \\
    8. Run strict validation on detector positives; reject reports whose relation, evidence, or impact does not survive source-level falsification. \\
    9. Deduplicate validated reports by root cause and likely fix; return root-level findings $F$. \\
    \bottomrule
  \end{tabular}
\end{table*}

\subsection{Business Path Extraction}

\tool first constructs a project-level call graph from Solidity source code. The graph includes public and external entries, internal helper calls, and cross-contract calls that can be resolved statically. The unit of analysis is not limited to one function. We include single-function paths, entry-to-leaf chains, and branch segments around function-call forks because each of these can represent a business operation boundary.

To avoid an unbounded path explosion, the current prototype focuses on externally reachable business paths and relation-relevant segments. A straight-line call chain without function-call branching is retained as one path. When a path branches into multiple function-call successors, \tool retains the fork segment and each reachable branch path because relation violations often occur at the point where one branch applies a guard, fee, or state update that another branch omits. This design preserves middle segments without enumerating every arbitrary subpath.

\subsection{Path Fact Summarization}

For each path, \tool extracts a compact fact summary. Static facts include function visibility, mutability, modifiers, require-like checks, storage reads and writes, mapping keys when recoverable, ETH/token transfers, external calls, events, custom errors, and call targets. The summary is intentionally lossy but relation-oriented: it keeps facts that can support obligations across paired paths.

View functions are handled selectively. A pure query that is not called by a state-modifying or state-dependent path is usually not relation-relevant. A view path is retained when it predicts execution behavior, participates in price/fee computation, or is called by a state-modifying path. This rule keeps preview/execute and helper/core relations while removing unrelated read-only utilities.

\subsection{Candidate Pair Reduction}

Given $n$ retained paths, exhaustive semantic comparison requires $O(n^2)$ pair judgments. \tool reduces this space with a static ranker before invoking an LLM. The ranker scores two paths using shared storage footprint, overlapping external calls, related entry names, common helper usage, actor/value-flow similarity, and relation-shape hints such as single/batch or wrapper/core naming. The ranker is not a final classifier; it only prioritizes pairs whose facts suggest a possible relational oracle.

The current prototype uses a conservative strategy: it prefers ranking and budgeted cutoffs over hard deletion when a signal is uncertain. This is important because chiral pairs may be opposite rather than syntactically similar. For example, buy/sell paths may have inverse state deltas and different names but still touch the same supply, balance, fee, and timestamp state.

The ranker is a weighted sum over interpretable features:
\begin{equation}
score(p,q)=\sum_i \lambda_i f_i(\Phi(p),\Phi(q)).
\end{equation}
Important features include shared write variables, read/write overlap, common external callees, common internal helpers, relation-name hints, actor/value-flow overlap, and lifecycle-object overlap. The weights in the prototype are heuristic rather than learned. This is acceptable for the current stage because the ranker is used to prioritize LLM work, not to make final vulnerability decisions. In a larger benchmark, these weights should be tuned against held-out projects and reported as an ablation.

\begin{table*}[t]
  \caption{Static ranking features. The ranker is designed for candidate prioritization, not final proof.}
  \label{tab:ranker}
  \centering
  \small
  \begin{tabular}{p{0.18\linewidth}p{0.37\linewidth}p{0.34\linewidth}}
    \toprule
    Feature group & Signal & Why it helps \\
    \midrule
    Storage overlap & Shared reads/writes, same mapping base, same object identifier & Related business operations usually touch the same protocol object or accounting state. \\
    Call overlap & Shared helpers, shared external calls, common settlement sink & Wrappers, dispatchers, and variants often converge on common implementation points. \\
    Relation names & claim/batchClaim, buy/sell, preview/execute, create/update & Naming is weak but useful when combined with state and call facts. \\
    Actor/value facts & Same payer/receiver/signature/refund variables or same ETH/token movement & Many chiral bugs are actor-binding or value-flow mismatches. \\
    Lifecycle facts & Same art, market, position, order, credential, or pool identifier & Lifecycle paths should preserve state-machine constraints over the same object. \\
    Branch/fork facts & Shared prefix with different call successors & Mismatches frequently occur at function-call branch boundaries. \\
    \bottomrule
  \end{tabular}
\end{table*}

\subsection{Semantic Filtering and Detection}

The LLM stage has two tasks. The semantic filter decides whether two paths plausibly form a chiral relation and assigns a relation type. It is asked to keep both similar and opposite relations, and to avoid rejecting a pair merely because path lengths differ. The detector then asks whether the relation induces a violated obligation with concrete source evidence and impact. The prompt emphasizes that a report should identify the relation, the expected obligation, the observed mismatch, the source evidence, and the impact thesis.

This division reduces cost compared with asking for full vulnerability analysis on every static pair. It also makes failures easier to inspect: a false negative can be attributed either to pair recall, semantic relation recognition, or vulnerability judgment. In our implementation, both stages are executed with a low-cost coding-agent model and can be parallelized across independent pairs.

The semantic filter is intentionally relation-first. It does not ask ``is there a bug?'' as the first question. Instead, it asks whether the pair has a plausible business correspondence and what type of correspondence it is. This matters because a model can miss inverse or wrapper relations when prompted only for similarity. The filter therefore explicitly allows pairs with different starts, different path lengths, different function names, and opposite economic directions.

The detector is obligation-first. Given a kept relation, it asks which obligation would be expected under that relation type and whether the source violates it. The detector must name the relation, the expected behavior, the observed divergence, the source evidence, and the impact thesis. This prompt structure is a lightweight way to force the model to expose its oracle rather than merely produce a vulnerability-sounding conclusion.

\subsection{Prompt Contracts}

We treat prompts as contracts rather than informal instructions. The semantic-filter contract requires four fields: relation verdict, relation type, short rationale, and uncertainty. The detection contract requires six fields: pair identity, relation type, violated obligation, evidence, impact, and confidence. The validation contract is falsification-first: it asks whether the relation is justified, whether the cited evidence exists, whether another code path invalidates the thesis, and whether the impact remains after considering protocol context.

This contract style gives the pipeline inspectable failure modes. If the semantic filter rejects a true pair, the failure is a relation-recognition miss. If the detector keeps a real relation but invents impact, the failure is an obligation-to-impact miss. If the strict validator rejects a true issue because it overlooks evidence, the failure is a validation false negative. These distinctions are useful for improving the system because each stage can be debugged independently.

\subsection{Validation and Root-Cause Deduplication}

The detector output is intentionally recall-oriented. \tool therefore applies a stricter validator that attempts to falsify each finding against the full source context. The validator checks whether the pair is real, whether cited code evidence exists, whether the expected relation is justified, whether an alternate code path invalidates the thesis, and whether the impact survives a skeptical reading.

Finally, \tool deduplicates findings at the root-cause level. Two reports are merged when they point to the same underlying defect and the same likely fix, even if they arise from different pair shapes or use different wording. This step is essential for usability: relation-based detection can discover the same root cause through many path pairs.

Root-cause deduplication is stricter than textual clustering. Two reports are not merged merely because they mention the same function. They are merged when a developer would likely apply the same code fix to address both reports. Conversely, reports in the same code area remain separate when they require different fixes. In the Phi study, for example, art initialization zero-address validation and art time-boundary validation live near each other but were split because they have different root causes and fixes.

\section{Implementation}

The current \tool prototype targets Solidity projects. Static extraction uses source-level parsing and call-graph reconstruction. LLM-based semantic filtering, detection, and strict validation are run through a coding-agent interface using a low-cost large language model. The validator prompt is falsification-first: it is instructed to reject findings when evidence is missing, when the relation is not justified, or when the claimed impact does not survive source-level reasoning.

\tool records token usage, wall-clock time, model cost estimates, intermediate pair decisions, detector outputs, validator verdicts, and deduplication clusters. These artifacts are used both for debugging the pipeline and for evaluating cost/recall tradeoffs. The implementation currently supports path-level analysis, semantic pair judgment, detection, strict validation, and manual root-cause review; full automated proof generation is left to future work.

\subsection{Artifact Design}

Every stage writes explicit artifacts. The static stage records retained paths, path summaries, and ranked pairs. The semantic stage records keep/maybe/reject decisions and relation rationales. The detection stage records per-pair vulnerability theses. The validation stage records valid, partially valid, discussion, and invalid decisions. The deduplication stage records cluster membership, representative reports, and manual root-cause notes.

This artifact design is important for a workflow that uses LLMs. Without intermediate artifacts, a failed run is difficult to interpret: it is unclear whether the system missed a pair, rejected a relation, failed to notice an obligation, hallucinated impact, or over-deduplicated reports. With artifacts, the pipeline can be audited stage by stage. The same artifacts also support later benchmark construction because they expose candidate-level and root-level labels.

\subsection{Parallel Execution}

The expensive stages are pair-local and can be parallelized. Semantic filtering can run independently for each static pair, detection can run independently for each kept pair, and validation can run independently for each finding group. The current implementation records per-item elapsed time and token usage so that cost estimates can be extrapolated to larger projects. Parallelism reduces wall-clock time, but it does not reduce model cost; therefore pair reduction remains central.

\subsection{Reproducibility Notes}

The current draft reports preliminary numbers from one full Phi run. A fully reproducible release should include the exact source revision, path extraction configuration, static ranking weights, prompt templates, model identifiers, retry policy, and all intermediate JSON artifacts. For arXiv release, we plan to include anonymized or sanitized artifacts when source licensing and disclosure constraints allow it. Until then, the paper treats the Phi results as a case study rather than a public benchmark.

\section{Preliminary Evaluation}

\subsection{Research Questions}

Our preliminary evaluation asks three questions:
\begin{itemize}
  \item \textbf{RQ1: Pair reduction.} Can static ranking and semantic filtering reduce the path-pair search space to a manageable set without removing relation-rich pairs?
  \item \textbf{RQ2: Detection utility.} Does relation-aware detection produce concrete smart-contract findings that are not merely wording variants of local syntactic rules?
  \item \textbf{RQ3: Cost and triage.} What are the cost and deduplication characteristics of an LLM-assisted chiral-analysis pipeline?
  \item \textbf{RQ4: Validation behavior.} How much does a strict falsification-first validator reduce noisy detector output, and what kinds of issues survive?
\end{itemize}

\subsection{Subject Protocol}

We evaluate \tool on the Phi protocol as a preliminary case study. Phi is useful for early evaluation because it contains multiple relation shapes: claim and batch-claim flows, adapter and factory flows, quote and execution paths, single and batch trading paths, lifecycle operations for art creation and update, and shared fee/reward accounting. We use the complete source available to the analyzer and manually inspect strict-positive findings at the root-cause level.

The protocol is not selected as a convenience example for a single known bug. It is useful because it stresses the exact relation shapes that motivated the model. It has cross-contract adapters, direct factory paths, state-changing trading paths, view pricing helpers, batched entry points, signature and Merkle claim variants, and lifecycle operations over art objects. These features allow us to observe whether a relation-based pipeline finds multiple classes of inconsistencies rather than a single hard-coded case.

The manual inspection protocol groups reports by root cause and likely fix. We intentionally do not count every pair-level report as a separate bug because relation analysis can rediscover one root cause through many pair shapes. We also keep partially valid reports separate from fully valid reports during validation, because early research prototypes benefit from preserving uncertain but plausible signals for later human triage.

\subsection{Main Results}

Table~\ref{tab:phi-results} summarizes the Phi run. Starting from 3,217 statically ranked candidate pairs, semantic filtering kept or marked 1,643 pairs for detection. The detector produced 201 cleaned finding/suspicious candidates. A local deduplication and validation pass reduced these to 101 finding groups. A strict validator retained 44 groups as valid or partially valid, and manual root-cause deduplication collapsed those groups to 13 effective unique issues.

\begin{table}[t]
  \caption{Preliminary Phi results. Counts are artifacts from one full \tool run.}
  \label{tab:phi-results}
  \centering
  \small
  \begin{tabular}{lr}
    \toprule
    Stage & Count \\
    \midrule
    Static candidate pairs & 3,217 \\
    Semantic keep/maybe pairs & 1,643 \\
    Detector no-issue pairs & 953 \\
    Detector not-pair pairs & 446 \\
    Detector suspicious candidates & 141 \\
    Detector finding candidates & 103 \\
    Cleaned finding/suspicious candidates & 201 \\
    Local deduplicated groups & 101 \\
    Strict-validator positives & 44 \\
    Manual root-cause issues & 13 \\
    \bottomrule
  \end{tabular}
\end{table}

The funnel shows two things. First, pair-level output is much noisier than root-level output. The 201 cleaned finding/suspicious candidates collapse to 101 local groups, then to 44 strict-positive groups, and finally to 13 root-level issues. Second, this reduction is not a sign that the detector is useless; it is a sign that relation-based discovery is high recall and duplicate-heavy. The same root cause may be visible through direct/batch, wrapper/core, and shared-settlement relations. A practical tool must therefore report both the final root issue and the supporting duplicate paths.

Table~\ref{tab:issue-types} summarizes representative root-cause issues from the 13 manually deduplicated strict-positive groups. The results cover multiple relation types, including preview/execute, single/batch, wrapper/core, lifecycle, and similar-flow relations.

\begin{table*}[t]
  \caption{Representative manually deduplicated issues from the Phi case study.}
  \label{tab:issue-types}
  \centering
  \small
  \begin{tabular}{p{0.15\linewidth}p{0.43\linewidth}p{0.18\linewidth}p{0.13\linewidth}}
    \toprule
    ID & Root-cause summary & Relation type & Manual priority \\
    \midrule
    M10 & Merkle proof does not bind art identifier, allowing proof reuse across related art assumptions & Similar flow / guard & High \\
    M13 & Art creation fee is documented or validated as basis points but consumed as wei & Value/unit & High \\
    M12 & Public tracking helpers can corrupt curator position state & State / access & Medium--high \\
    M05 & Single trade path lacks non-reentrancy and differs from batch trade ordering/guard behavior & Single/batch & Medium--high \\
    M02 & Claim and batch-claim paths handle overpayment/refund propagation inconsistently & Single/batch value flow & Medium \\
    M04 & Claimable adapter forwards raw value and can receive refunds instead of the original user & Wrapper/core value flow & Medium \\
    M09 & Signature-claim path relies on off-chain discipline where Merkle path enforces verification type on chain & Similar flow guard & Medium \\
    M01 & Creator-fee quote path and execution-related pricing helper disagree on zero-supply boundary behavior & Preview/execute & Low--medium \\
    \bottomrule
  \end{tabular}
\end{table*}

\subsection{Qualitative Finding Analysis}

The highest-priority Phi findings illustrate different parts of the taxonomy. The Merkle proof issue is a similar-flow guard violation: a proof should bind the business object being claimed, and the paired claim paths reveal that the eligibility proof does not sufficiently bind the art identifier. The art creation fee issue is a value-unit violation: the same configuration value is interpreted under incompatible unit assumptions. The public tracking-helper issue is a state/access violation: helper functions that should behave as internal accounting transitions are exposed as public state-mutating paths. The refund propagation issues are wrapper/core and single/batch value-flow violations: equivalent user-facing operations do not preserve the refund recipient.

Some retained issues are lower severity but still useful. Quote/execution mismatches may not directly steal funds if execution uses a safer helper, but they can mislead integrators and users. Delayed validation in a batch path may not bypass the final check, but it can reveal an inconsistent failure model or unnecessary gas griefing surface. Chiral analysis is useful partly because it exposes these inconsistencies with their relation context, allowing auditors to decide whether they are security findings, correctness bugs, or design discussions.

\subsection{Validator Effect}

The strict validator materially changes the output distribution. Before strict validation, the pipeline had 101 locally deduplicated groups. After strict validation, 44 groups remained valid or partially valid, and 57 were rejected. Manual review of strict invalids found candidate-level false kills and unstable duplicate judgments, but no clear high-confidence root-level miss in the reviewed Phi run. This suggests that a strict validator is valuable as a precision layer, but it should not be treated as an infallible oracle.

The validator's main benefit is not only rejecting false positives. It also forces the report format to become sharper. A detector may say that two paths are inconsistent; the validator asks whether the relation is justified, whether the evidence exists in the source, whether there is an alternate execution path that restores the invariant, and whether the impact is still meaningful. This falsification-first posture is especially important for LLM-assisted analysis because vulnerability-sounding text can otherwise hide weak evidence.

\subsection{Cost}

The full Phi run cost an estimated 212.70 CNY including semantic filtering, pair detection, reruns for missing items, and strict validation. The pre-strict sequential wall-clock time was approximately 4.01 hours, while strict validation consumed 10,820 seconds of summed per-item runtime across 101 item runs. These numbers are preliminary and depend on model pricing, prompt length, and parallelism. However, they show that the dominant cost is not the static graph analysis but the number of LLM pair judgments and validation runs.

\begin{table}[t]
  \caption{Cost breakdown for the Phi preliminary run.}
  \label{tab:cost}
  \centering
  \small
  \begin{tabular}{lrr}
    \toprule
    Stage & Time & Cost (CNY) \\
    \midrule
    Semantic filter & 5,539 s & 65.17 \\
    Pair detection & 7,664 s & 96.41 \\
    Missing reruns & 1,235 s & 14.34 \\
    Strict validation & 10,820 s & 36.79 \\
    \midrule
    Total & -- & 212.70 \\
    \bottomrule
  \end{tabular}
\end{table}

The cost table should be read as an engineering measurement rather than a fundamental limit. The cost can be reduced by improving static ranking, batching short relation judgments, caching source summaries, using smaller models for semantic filtering, and reserving stronger models for strict validation. Conversely, cost can increase on projects with more externally reachable paths or more cross-contract wrappers. This is why the paper reports the funnel and not only the final finding count.

\subsection{Interpretation}

The case study supports the central motivation of chiral analysis: many findings are easiest to express as violated path relations. Refund propagation issues arise by comparing wrapper or batch paths against direct core paths. Fee-unit and quote mismatches arise by comparing semantic expectations across related paths. Single/batch guard asymmetries arise because one entry point receives a hardening check that its paired path lacks. These examples are awkward to encode as isolated single-function detectors but natural under relation-induced obligations.

The evaluation is still preliminary. We have not yet built a large benchmark across many protocols, and the current ground truth relies on manual root-cause review rather than project-maintainer confirmation. We therefore present the Phi results as evidence that the modeling and pipeline are viable, not as a final measurement of general recall or precision.

\subsection{Planned Baselines and Ablations}

An arXiv version should make clear what remains to be evaluated. We plan to compare \tool against four baselines. The first is direct-agent auditing, where a coding agent reads the project in chunks and reports vulnerabilities without pair scaffolding. The second is static-only chiral ranking, where top-ranked pairs are manually or mechanically inspected without LLM semantic filtering. The third is embedding-only retrieval over functions or paths. The fourth is local detector output from existing smart-contract analyzers where applicable.

The key ablations are also clear from the pipeline. Removing branch-segment retention tests whether middle-path relations matter. Removing static ranking tests whether semantic filtering alone is affordable. Removing relation-type prompts tests whether the model loses inverse and wrapper pairs when asked only for similarity. Removing strict validation tests how much detector output is noise. Removing root-cause deduplication tests whether pair-level reporting is usable for auditors.

\section{Related Work}

\subsection{Smart-Contract Analysis}

Smart-contract analyzers commonly translate source or bytecode into an intermediate representation and check vulnerability patterns or semantic facts. Securify extracts compliance and violation patterns from smart contracts~\cite{tsankov2018securify}. SmartCheck converts Solidity programs into an XML-like representation and applies pattern queries~\cite{tikhomirov2018smartcheck}. Slither builds a Solidity static-analysis framework around an intermediate representation and detector infrastructure~\cite{feist2019slither}. \tool is complementary to these systems: it uses static facts and call-graph structure, but its core oracle is a relation between business paths rather than a local vulnerability template.

\subsection{Rule Mining and Inconsistency Detection}

Prior SE work has shown that bugs can be detected by mining implicit rules or by finding inconsistent behavior across related code. Engler et al. detect deviant behavior by inferring common programming beliefs and flagging violations~\cite{engler2001bugs}. PR-Miner mines implicit programming rules from large codebases~\cite{li2005prminer}. CP-Miner and later inconsistency detectors identify suspicious differences across cloned or related code regions~\cite{li2006cpminer,ahmadi2021fics}. Chiral analysis follows the same broad principle that inconsistency can be a bug oracle, but it differs in the matching unit: the corresponding objects are business paths that may be syntactically different, cross-contract, or semantically inverse.

\subsection{Metamorphic Testing}

Metamorphic testing addresses the oracle problem by defining relations between multiple executions instead of requiring an exact expected output for one execution~\cite{segura2016metamorphic}. Chiral analysis can be viewed as a static, contract-specific analogue of this idea. A chiral relation maps one business path to another and induces expected relations over state, outputs, and observable side effects. Unlike conventional metamorphic testing, our current prototype does not require generating and executing concrete test inputs for every relation; it approximates relation violations using static facts and LLM-assisted semantic reasoning.

\subsection{LLM-Assisted Program Analysis}

Large language models are increasingly used to summarize code, inspect security reports, and support agentic debugging. \tool uses LLMs in a constrained role: they judge relation semantics and vulnerability theses over a reduced candidate set, while static extraction, pair ranking, artifact tracking, validation, and deduplication structure the workflow. This design treats the LLM as a semantic component inside a program-analysis pipeline rather than as a one-shot auditor over an entire codebase.

\subsection{Positioning Against Direct Agent Auditing}

Direct agent auditing is a strong practical baseline because modern coding agents can read source files, follow call chains, and reason about protocol logic. However, direct auditing has weak observability: when it misses a bug, it is difficult to know whether the model failed to inspect the right files, failed to connect related paths, failed to notice the mismatch, or failed to classify the impact. Chiral analysis makes this process more inspectable by externalizing path pairs, relation labels, obligation checks, and validator decisions.

The claim is not that pair-first analysis is always more powerful than an unrestricted agent. The claim is that pair-first analysis gives the agent a smaller and more testable semantic task. It also creates reusable artifacts for benchmark construction. If a direct agent finds a bug, the result is useful; if a chiral pipeline finds or misses a pair, the intermediate artifacts explain where the pipeline succeeded or failed.

\section{Discussion}

\subsection{Why Pair First?}

One might ask why a coding agent cannot simply read the code in chunks and directly report all vulnerabilities. The pair-first design gives the model a narrower and more auditable task. Instead of asking for an open-ended audit over a large project, \tool asks whether two concrete paths form a relation and whether that relation is violated. This reduces prompt ambiguity, records intermediate decisions, and makes cost proportional to a ranked pair set rather than uncontrolled source exploration.

This design also matches how many human auditors reason. Auditors often notice that one flow has a check and another related flow does not, or that a batch path uses a slightly different settlement order than the single path. Chiral analysis attempts to make that comparison explicit and repeatable. The benefit is not only automation; it is a vocabulary for describing why a business-logic inconsistency matters.

\subsection{Precision and Recall Tradeoff}

Chiral analysis is naturally recall-oriented at the candidate stage. If the static ranker removes a true pair, later semantic reasoning cannot recover it. For this reason, the current prototype prefers conservative retention before semantic filtering and uses strict validation only after detection. This design can produce duplicate and noisy candidates, but root-cause deduplication makes the final output manageable.

\subsection{Limitations}

The current implementation has several limitations. First, path extraction is an approximation of Solidity semantics and may miss dynamically resolved calls or complex proxy behavior. Second, the LLM stages can produce both false positives and false negatives, especially when source context is incomplete. Third, the current evaluation is a preliminary case study rather than a large-scale benchmark. Fourth, the system reports relation violations but does not yet automatically construct executable proof-of-concept transactions.

\subsection{Where Chiral Analysis May Fail}

Chiral analysis can miss vulnerabilities that do not have a meaningful paired path. A single isolated function with a unique bug may be better handled by a conventional detector or a direct audit. The method can also miss pairs when the static ranker fails to connect them, especially if they share little storage footprint and are linked only by high-level documentation. Conversely, it can produce noise when two paths touch the same state for legitimate but unrelated reasons.

The approach also depends on model judgment for semantic relation recognition. A model may reject an inverse relation because the names differ, or it may overstate a mismatch because it ignores protocol policy. The strict validator reduces this risk but does not eliminate it. A future implementation should combine model judgment with more symbolic path conditions, stronger source slicing, and project documentation when available.

\section{Future Work}

\subsection{Benchmark Construction}

The next major step is to build a benchmark with root-level labels. A useful benchmark should include protocols with direct/batch flows, router/core flows, preview/execute paths, inverse financial operations, lifecycle state machines, and shared settlement helpers. Each benchmark item should record the relation pair, the expected obligation, the source evidence, the root cause, and the final label. Pair-level labels alone are insufficient because one root cause can appear in many pair-level candidates.

\subsection{Witness Generation}

The current system reports source-level evidence and impact theses, but it does not automatically construct executable proofs. A stronger version should attempt witness generation for selected high-confidence findings. For value-flow issues, this may involve symbolic or concrete transaction construction. For guard asymmetries, it may involve generating two related calls where one path accepts an input that the paired path rejects. For preview/execute mismatches, it may involve generating states where the quote diverges from execution.

\subsection{Learning the Ranker}

The current ranker uses heuristic weights. A larger labeled dataset would allow learning a ranking model over path-pair features. The goal would not be to replace semantic detection, but to reduce the number of LLM calls needed to achieve a target recall. A learned ranker could also expose which static features are most predictive for each relation type.

\section{Threats to Validity}

\textbf{Internal validity.} Our manual deduplication and issue-priority labels may contain judgment errors. To reduce this risk, strict validation attempts to falsify each candidate against source evidence, and manual grouping merges findings only when they share the same root cause and likely fix.

\textbf{External validity.} The preliminary evaluation uses one protocol. Phi contains several useful relation shapes, but it may not represent all Solidity projects. A broader evaluation should include DeFi protocols, NFT protocols, governance systems, bridges, and vault/router architectures.

\textbf{Construct validity.} Chiral vulnerability is a broader concept than standard vulnerability templates. Some relation violations may be low-severity inconsistencies rather than exploitable bugs. We therefore distinguish suspicious candidates, validated vulnerabilities, and root-level findings.

\textbf{Cost validity.} LLM cost depends on model pricing, prompt size, parallelism, caching, and retry policy. Reported cost should be interpreted as an implementation measurement, not a fixed property of the method.

\textbf{Disclosure validity.} Some findings may require project-specific disclosure or maintainer confirmation before being treated as confirmed vulnerabilities. This draft reports them as manually triaged research findings rather than externally confirmed advisories.

\section{Conclusion}

This paper introduced chiral analysis, a relational approach to smart-contract vulnerability detection. The central idea is to treat corresponding business paths as implicit specifications for each other and to detect violations of the obligations induced by their relation. We formalized chiral relations, defined obligation dimensions and vulnerability witnesses, and implemented a prototype pipeline that combines static path analysis with LLM-assisted semantic detection and strict validation. A preliminary Phi case study suggests that the approach can uncover meaningful business-logic inconsistencies across single/batch, wrapper/core, preview/execution, and similar-flow relations. Future work will expand the benchmark, compare against direct-agent and static-analysis baselines, and strengthen witness generation.

\bibliographystyle{plain}
\bibliography{references}

\end{document}